\documentclass[onecolumn,floatfix,showpacs,nofootinbib,11pt,groupedaddress,superscriptaddress,]{revtex4-1}
\usepackage{amssymb,amsmath,graphicx}
\usepackage{color}
\usepackage[colorlinks,hyperindex]{hyperref}
\usepackage{slashed}
\hypersetup{hidelinks,backref=true,pagebackref=true,hyperindex=true,colorlinks=true,breaklinks=true,urlcolor= blue}
\hypersetup{%
  colorlinks = true,
  linkcolor  = blue
}
\usepackage{bookmark}
\usepackage{hyperref,color,xcolor}
\hypersetup{hidelinks,backref=true,pagebackref=true,hyperindex=true,colorlinks=true,breaklinks=true,urlcolor= blue}
\hypersetup{%
  colorlinks = true,
  linkcolor  = blue
}
\usepackage{makeidx,upgreek}
\newcommand{\beq}{\begin{eqnarray}}
\newcommand{\benu}{\begin{enumerate}}
\newcommand{\enu}{\end{enumerate}}
\newcommand{\eeq}{\end{eqnarray}}
\newcommand{\be}{\begin{equation}}
\newcommand{\ee}{\end{equation}}

\newcommand{\ba}{\begin{eqnarray}}
\newcommand{\ea}{\end{eqnarray}}
\begin{document}
\title{Unveiling a spinor field classification with non-Abelian gauge symmetries}
\author{Luca Fabbri}
\email{fabbri@dime.unige.it}
\affiliation{DIME, Universit\`a di Genova, P. Kennedy Pad. D, 16129 Genova, Italy}
\author{Rold\~ao~da~Rocha}
\email{roldao.rocha@ufabc.edu.br}
\affiliation{CMCC, Universidade Federal do ABC, 09210-580,
Santo Andr\'e, Brazil.}
\begin{abstract}
A spinor fields classification with non-Abelian gauge symmetries is introduced, generalizing the 
the U(1) gauge symmetries-based Lounesto's classification. Here, a more general classification, contrary to the Lounesto's one, encompasses spinor multiplets, corresponding to non-Abelian gauge fields. \color{black}{The particular case of SU(2) gauge symmetry, encompassing electroweak and electromagnetic conserved charges,
 is then implemented by a non-Abelian spinor classification, now involving 14 mixed classes of spinor doublets.
A richer flagpole, dipole, and flag-dipole structure naturally descends from this general classification. The Lounesto's classification of spinors is shown to arise as a Pauli's singlet, into this more general classification. }

\end{abstract}
\maketitle
\section{Introduction}
The Lounesto's spinor fields classification \cite{lou2} represents an assortment of all spinor field in Minkowski spacetime that has been shown to be complementary to the Cartan's and the Wigner's  classifications of spinors. In the Lounesto's spinor fields classification the standard Majorana, Weyl, and Dirac spinor fields are representatives of very particular subsets in different classes of spinors, classified according to their bilinear covariants. Several non-standard spinors, charged and neutral as well,  have been studied. Refs. \cite{daSilva:2012wp,2017} encode an up-to-date on the Lounesto's classification. Besides, concrete examples of non-standard spinor fields were provided in, e. g., \cite{exotic,esk,VazJr:2016jnp}. This classification has been extended, in order to further encompass new classes of spinors on higher dimensional spacetimes. For example, new spinors were constructed  on 7d manifolds, that in particular arise as new solutions of the Euler-Lagrange equations in the AdS$_4\times S^7$ compactifications in string theory \cite{BBR}, as well as for  AdS$_5 \times S^5$ compactifications \cite{deBrito:2016qzl}.

Both from the formal and the pragmatic points of view, 
the Lounesto's classification of spinors is well established and successful, for its huge variety 
of applications and exploratory features on the search of new fermions fields \cite{daSilva:2012wp}. However, it is remarkably limited in the context of gauge symmetries, just holding for the case of  U(1) (abelian) gauge symmetries.  In fact, spinors in the Lounesto's classification can be split into classes of charged and neutral spinors, under the conserved electric charge that is evinced 
from the Noether's theorem, due to the U(1) gauge symmetry underlying the equations for motion ruling all the spinor fields. Therefore, the Lounesto's classification is not able to encompass spinor multiplets, corresponding to non-Abelian gauge fields. In particular, it does not encode electroweak and strong conserved charges. In fact, the Standard Model (SM) of elementary particles is described by a gauge theory, effectively governed by the gauge group SU(3) $\times$ SU(2) $\times$ U(1), 
describing strong, weak and electromagnetic interactions. Those interactions are implemented  when the corresponding
bosonic gauge fields, that include 8 massless gluons, 3 massive bosons, $W_\pm$ and $Z$,  and 1  photon, are exchanged, to respectively describe strong and electroweak. The fermionic sectors of the theory describe matter and it is encoded into 3-fold families of quarks and leptons, together with their antiparticles. The Lounesto's classification can solely  encompass Abelian gauge symmetries, with conserved electric charge.
 Exploring algebraic solutions for the U(1) electromagnetic potential appearing in the Dirac equation,  Refs. \cite{Booth:2001ey,Fabbri:2017pwp} showed the prototypical inversion theorem for the real vector potential. The SU(2)  case was scrutinized in Ref. \cite{Inglis:2012rg}. 
 
The main aim here is to propose a spinor field classification that further encompasses  non-Abelian gauge symmetries, encoding the  Lounesto's classification of spinors that correspond, into this extended classification, to a Pauli's singlet, also encompassing electroweak and electromagnetic conserved charges corresponding to the SU(2) $\times$ U(1) symmetry. The extended, non-Abelian, Fierz identities are then here scrutinized.

This paper is organized as follows: after reviewing the Lounesto's classification, the Fierz identities and the Fierz aggregate in Sect. II, Sect. III is devoted to analyze the inversion of the Dirac equation, showing that it holds for the case of type-1 regular spinors.  The electromagnetic potential can be also expressed as spinor fields for the case of type-2 and type-3 regular spinors,  and we show that the inversion can not be implemented for singular spinors. In Sect. IV a non-Abelian spinor classification is implemented, with a richer flagpole, dipole, and flag-dipole structure, {\color{black}{where 14 classes of spinor doublets are allowed, corresponding to bispinor fields}}. The particular case of SU(2) gauge symmetry is implemented, making the Lounesto's classification of spinors to correspond to a Pauli's singlet in this quite more general classification that encompass electroweak and electromagnetic conserved charges. In Sect. V, non-Abelian Fierz aggregates and generalized Fierz identities, corresponding to the proposed doublet spinors, 
are listed and rederived. 
\section{The U(1) classification of spinor fields, the Fierz aggregate and the Fierz identities}

(Classical) spinor fields are objects defined on  Minkowski spacetime $M$ that are well known to carry the
${\left(\frac12,0\right)}\oplus{\left(0,\frac12\right)}$ Lorentz group representations.   With respect to an arbitrary basis $\{\upgamma^\mu\}\subset \sec\Omega(M)$, where $\Omega(M)=\oplus_{i=0}^4\Omega^i(M)$ denotes the exterior bundle, 
the bilinear covariants are  sections  of $\Omega(M)$, whose splitting is then represented by   \cite{Cra}
\begin{center}
\begin{tabular}[c]{||c|| c|| c|| c|| c||}
\hline\hline
$\;\;\sec\Omega^0(M)\;\;$&$\;\;\sec\Omega^1(M)\;\;$&$\;\;\sec\Omega^2(M)\;\;$&$\;\;\sec\Omega^3(M)\;\;$&$\;\;\sec\Omega^4(M)\;\;$
\\\hline
 $\sigma=\bar{\psi}\psi$ & $\mathbf{J}=J_{\mu}\upgamma^{\mu}$ & $\;\;\mathbf{S}=S_{\mu\nu }\upgamma^{\mu}\wedge \upgamma^{ \nu }\;\;$&$\mathbf{K}= K_{\mu }\upgamma^{\nu }$&$\omega=\bar{\psi}\upgamma_{5}\psi$\\
\hline\hline
\end{tabular}
\end{center} 
where $J_{\mu}=\bar{\psi}\upgamma _{\mu }\psi,$ $
S_{\mu\nu}=\bar{\psi}[\upgamma _{\mu},\upgamma_{
\nu }]\psi,$ and $K_{\mu }=\bar{\psi}\upgamma_{5}\upgamma _{\mu }\psi$ are the respective coefficients of the Lorentz bilinear covariants, in the above table. Also, $\upgamma_5=i\upgamma_0\upgamma_1\upgamma_2\upgamma_3$ is the chiral operator implemented by the volume element (for the Clifford product denoted by juxtaposition); the Dirac-conjugated spinor reads $\bar\psi=\psi^\dagger\upgamma_0$, and hereon $\upgamma_{\mu\nu}:=\frac{i}{2}[\upgamma_\mu, \upgamma_\nu]$. Besides,   $\upgamma_{\mu }\upgamma _{\nu
}+\upgamma _{\nu }\upgamma_{\mu }=2\eta_{\mu \nu }\mathbf{1}$, {where $\eta_{\mu\nu}$ denotes the 
Minkowski metric.}
The physical observables, exclusively for the Dirac's  theory describing the electron, are realized by the bilinear covariants. In fact, the 1-form current density $\mathbf{J}$, the 2-form spin density $\mathbf{S}$, and the 1-form chiral current density $\mathbf{K}$, satisfy, together to the scalar and pseudoscalar bilinears, the Fierz identities \cite{lou2} 
\begin{subequations}
\begin{eqnarray}\label{fifi}
-\omega{S}_{\mu\nu}+\sigma\epsilon_{\mu\nu}^{\;\;\;\alpha\beta}S_{\alpha\beta}&=&\epsilon_{\mu\nu\alpha\beta}J^\alpha {K}^\beta,\\\eta_{\mu\nu}J^\mu J^\nu+\eta_{\mu\nu}K^\mu K^\nu&=&0=\eta_{\mu\nu}J^\mu K^\nu,\\ \eta_{\mu\nu}J^\mu J^\nu&=&\omega^{2}+\sigma^{2}\,.  
\end{eqnarray}
\end{subequations}
\noindent  

The Lounesto's classification reads \cite{lou2}:
\begin{subequations}
\begin{eqnarray}
&&(1)\;\;\;\mathbf{K}\neq 0, \;\;\;\mathbf{S}\neq0,\;\;\;\omega\neq0,\;\;\;  \sigma\neq0,\;\;\text{}
\label{tipo1}\\
&&(2)\;\;\;\mathbf{K}\neq 0, \;\;\;\mathbf{S}\neq0,\;\;\;\omega=0,\;\;\;  \sigma\neq0,\;\;\label{tipo2}\\   
&&(3)\;\;\;\mathbf{K}\neq 0, \;\;\;\mathbf{S}\neq0,\;\;\;\omega\neq0,\;\;\;  \sigma=0,\label{tipo3}\\  
&&(4)\;\;\;\mathbf{K}\neq 0, \;\;\;\mathbf{S}\neq0,\;\;\;\omega=0=\sigma,  \;\;\text{}\quad\qquad\label{tipo4}\\
&&(5) \;\;\;\mathbf{K}= 0, \;\;\;\mathbf{S}\neq0,\;\;\;\omega=0=\sigma,\text{}\quad\qquad\label{tipo5}\\
&&(6)\;\;\;\mathbf{K}\neq 0, \;\;\;\mathbf{S}=0,\;\;\;\omega=0=\sigma.\;\;\;\text{}\quad\qquad\label{tipo6}
\end{eqnarray}
\end{subequations}
The condition $\mathbf{J}\neq0$ holds for all spinors into the above classes (\ref{tipo1} -- \ref{tipo6}).
Other classes corresponding to $\mathbf{J}=0$ have been derived in Ref. \cite{EPJC}, whose representative spinors have been conjectured to be ghost spinors.  The  most general representative  spinor fields of  each
Lounesto's spinor class, were listed in Ref. \cite{Cavalcanti:2014wia}. Moreover, a gauge spinor field  classification have also been proposed in Ref. \cite{Fabbri:2016msm}.

Singular spinors consist of flag-dipole, flagpole, and dipole spinors, respectively in the fourth, fifth, and sixth classes in the just mentioned six classes (\ref{tipo1} -- \ref{tipo6}). The standard Dirac spinor is an element of the set of regular spinors in class 1. Moreover, Majorana spinors are neutral spinors that embrace particular realizations of flagpole type-5 spinors. The chiral Weyl spinors consist of a tiny subset of dipole spinors. In fact, in Ref. \cite{Cavalcanti:2014wia} one  sees that chiral spinors are in the classes 6 that consists of dipole spinors, however only chiral spinors that satisfy the Weyl equation are Weyl spinors.
Since type-5 spinors phenomenologically accommodate mass dimension one spinors \cite{daRocha:2005ti,Alves:2014kta}, the class 6 might also accommodate 
mass dimension one spinors, whose dynamics, of course, is not ruled by the Weyl equation. Nevertheless, the classes  (\ref{tipo1} -- \ref{tipo6}) provide a comprehensive sort of new possibilities that have not been explored yet \cite{esk}.

 The Fierz identities (\ref{fifi}) do not hold for singular spinors. Based on a Fierz aggregate, 
 \begin{subequations}
 \begin{eqnarray}
\mathbf{Z}= (\omega-\mathbf{K})\upgamma_{5}+i\mathbf{S}+ \mathbf{J} +\sigma \,, \label{Z}\label{zsigma}
\end{eqnarray}
 the Fierz identities (\ref{fifi}) can be replaced  by the  most general equations 
\begin{eqnarray}
&&4i\omega\mathbf{Z}= -\mathbf{Z}\upgamma_{5}\mathbf{Z},\\
&&4iJ_{\mu}\mathbf{Z}= -\mathbf{Z}\upgamma_{\mu}\mathbf{Z},\\
&&4iS_{\mu\nu}\mathbf{Z}=-\mathbf{Z}\upgamma_{\mu}\upgamma_{\nu}\mathbf{Z},\\
&&4iK_{\mu}\mathbf{Z}=-\mathbf{Z}\upgamma_{5}\upgamma_{\mu}\mathbf{Z}\,.
\end{eqnarray}
\end{subequations} The above equations are reduced to Eqs. (\ref{fifi}), in the case where 
both $\sigma$ and $\omega$ are not equal zero, e. g., for type-1 spinor regular spinor fields in the (\ref{tipo1}) Lounesto's class. When $\upgamma^0{\bf Z}^\dagger \gamma^0 = {\bf Z}$, then the Fierz aggregate is a self-conjugated structure called a boomerang \cite{lou2}.

 The 1-form field {\bf J} is interpreted as being a pole, and flagpoles are consequently elements of the class 5 in Lounesto classification. In fact, for this one has ${\bf K} = 0$ and ${\bf S}\neq 0$, being the flagpole hence characterized by the non-vanishing  ${\bf S}$ and ${\bf K}$. Besides, as type-4 spinors have the 2-form field ${\bf S}$ and the 1-form fields {\bf J} and ${\bf K}$ non null, together they corresponding to a flag-dipole structure.  For type-6 spinors, {\bf J} and {\bf K} are the only bilinears that are not null and, then, they do correspond to a dipole  structure. 
The bilinear covariants also satisfy \cite{daRocha:2005ti}:
\begin{subequations}
\begin{align}
\eta^{\mu\alpha}{S}_{\mu\nu}{J}_{\alpha}\upgamma^\nu-\eta^{\nu\alpha}{S}_{\mu\nu}{J}_{\alpha}\upgamma^\mu   &  =\omega{K}_\rho\upgamma^\rho,\label{fie1}\\
\eta^{\mu\alpha}{S}_{\mu\nu}{K}_{\alpha}\upgamma^\nu-\eta^{\nu\alpha}{S}_{\mu\nu}{K}_{\alpha}\upgamma^\mu   &  =\omega{J}_\rho\upgamma^\rho,\\
i\epsilon_{\mu\nu}^{\;\;\;\rho\tau}(\eta^{\mu\alpha}{S}_{\rho\tau}{J}_{\alpha}\upgamma^\nu-\eta^{\nu\alpha}{S}_{\rho\alpha}{J}_{\alpha}\upgamma^\mu)   &  =2\sigma{K}_\rho\upgamma^\rho,\\
i\epsilon_{\mu\nu}^{\;\;\;\rho\alpha}(\eta^{\mu\alpha}{S}_{\rho\alpha}{K}_{\alpha}\upgamma^\nu-\eta^{\nu\alpha}{S}_{\rho\alpha}{K}_{\alpha}\upgamma^\mu)   &  =2\sigma{J}_\rho\upgamma^\rho,\\
S_{\mu\nu}S_{\rho\alpha}\eta^{\nu\rho}\eta^{\mu\alpha}&=-\omega^{2}+\sigma^{2},\\
i\epsilon_{\rho\alpha}^{\;\;\;\tau\xi}S_{\mu\nu}S_{\tau\xi}\eta^{\nu\rho}\eta^{\mu\alpha}&=-4\omega\sigma,\\
\eta^{\mu\alpha}{S}_{\mu\nu}{J}_{\alpha}\upgamma^\nu-\eta^{\nu\alpha}{S}_{\mu\nu}{J}_{\alpha}\upgamma^\mu +J_{\mu}S_{\nu\rho}\upgamma^\mu\wedge\upgamma^\nu\wedge\upgamma^\rho  &  =-\omega K_\tau\upgamma^\tau+\frac{i}{2}\sigma \epsilon_{\alpha\beta\tau\xi}K^\alpha\upgamma^\beta\wedge\upgamma^\tau\wedge\upgamma^\xi\\
\eta^{\mu\alpha}{S}_{\mu\nu}{K}_{\alpha}\upgamma^\nu-\eta^{\nu\alpha}{S}_{\mu\nu}{K}_{\alpha}\upgamma^\mu +K_{\mu}S_{\nu\rho}\upgamma^\mu\wedge\upgamma^\nu\wedge\upgamma^\rho  &  =-\omega J_\tau\upgamma^\tau+\frac{i}{2}\sigma \epsilon_{\alpha\beta\tau\xi}J^\alpha\upgamma^\beta\wedge\upgamma^\tau\wedge\upgamma^\xi\\
\epsilon_{\alpha\beta\mu\nu}S^{\alpha\beta}S^{\mu\nu}\upgamma_5 + \frac12S_{\mu\nu}S^{\mu\nu}&=\omega^{2}-\sigma
^{2}-2i\omega\sigma\upgamma_{5},
 \label{FIERZ}%
\end{align}
\end{subequations}

A spin-$\frac{1}{2}$ fermion, with charge $e$, is ruled by the Dirac equation $
(\slash\!\!\!\partial-e\slashed{A}(x)-m)\psi(x)=0$, with $\slashed{\partial} = \upgamma^\mu\partial_\mu$,  mass $m$,  and  electromagnetic potential $\slashed{A}(x)=\upgamma^{\mu}A_{\mu}(x)$. The current density $J^\mu$ is always conserved, related to the U(1) symmetry, $\partial_{\mu}J^{\mu}=0$, whereas the chiral current $
\partial_{\mu}K^{\mu}=-2i\, m\bar\psi\upgamma_{5}\psi$ is just conserved for $m=0$. In fact, the Dirac Lagrangian that originates this equation is U(1) invariant, namely, by the transformations $\psi(x)\mapsto e^{i\theta(x)}\psi(x)$ and $A_\mu(x)\mapsto A_\mu(x)+\frac{1}{e}\partial_\mu\theta(x)$. The U(1) covariant Dirac equation was shown to be equivalent to the following expressions for the inversion of the electromagnetic potential \cite{Booth:2001ey,Inglis:2012rg}:
\begin{eqnarray}
A_{\mu}&=&\frac{i}{2q\bar\psi\psi}\left[{\bar\psi\upgamma_{\mu}\slashed{\partial}\psi-\bar\psi\overleftarrow{\slashed{\partial}}\upgamma_{\mu}\psi-2mJ_{\mu}}{\bar\psi\psi}\right]\label{t1}\\
&=&\frac{i}{2q{\bar\psi\upgamma_{5}\psi}}\left[{\bar\psi\upgamma_{5}\upgamma_{\mu}\slashed{\partial}\psi+\bar\psi\overleftarrow{\slashed{\partial}}\upgamma_{5}\upgamma_{\mu}\psi}\right].\label{t2}
\end{eqnarray}
Eqs. (\ref{tipo1} -- \ref{tipo3}) show that the inversion (\ref{t1})
exists for spinors in Lounesto's classes 1 and 3, whereas the inversion (\ref{t2}) holds  for spinors in Lounesto's classes 1 and 2. For the other cases, including singular spinors, there is no inversion, in particular  for Weyl spinors, that satisfy $\bar\psi\psi = 0 = \bar\psi\upgamma_5\psi$.

\section{Non-abelian spinor classification}

Starting with the SU(2) gauge group, with associated Lie algebra $\mathfrak{su}(2)$ generated by the set  $\{\tau_{a}\}$ ($a=1,2,3$) of generators, satisfying  $[\frac{\tau_{a}}{2},\frac{\tau_{b}}{2}]=i\,\epsilon_{ab}{}^{c}\frac{\tau_{c}}{2}$,
non-Abelian SU(2) gauge fields $\slash\!\!\!\!{\mathbf{W}}=W_\mu\upgamma^\mu$ can be thought of as being a matrix of the type generated by an infinitesimal gauge transformation, meaning that the $W_\mu$  takes values in $\mathfrak{su}(2)$ and, therefore, can be split as 
$W_\mu= W_\mu^a\tau_a$, where the $W_{a\mu}$ are the SU(2)-Yang-Mills fields. The field strength is then given by $G_{\mu\nu}=\partial_\mu W_\nu - \partial_\nu W_\mu - [W_\mu, W_\nu]$.  
Requiring that  the Lagrangian for spinor fields must be  invariant under local SU(2) transformations,  the SU(2)  gauge covariant Dirac equation that governs a doublet spinor $\uppsi$, with SU(2) gauge field interactions,  reads 
\begin{equation}\label{diracsu2}
\left[i\slash\!\!\!\partial-\frac{g}{2}{\mathbf{\tau}}\cdot\slash\!\!\!\!{\mathbf{W}}-m\right]\uppsi=0,
\end{equation}
where $\tau_0={\rm id}_{2\times 2}$ and $g$ drives the Yang-Mills field running coupling.
 Remembering the definition of the charge conjugate spinor,  $
\uppsi^{{\rm c}}=C\bar\uppsi{}^\intercal=i\,\upgamma^{2}\upgamma^{0}\bar\uppsi{}^\intercal$,
 the complex conjugate of Eq. (\ref{diracsu2}), multiplied by $I\otimes U$, where $U$ is an operator that implements the parity and the complex conjugation,  $U\upgamma_{\mu}U^{-1}=-\upgamma_{\mu}^*$, such that $\uppsi^{{\rm c}}=U\uppsi^{*}$, yields \begin{eqnarray}
\left[\upgamma^{\mu}\left(i\,\partial_{\mu}+\frac{g}{2}{\mathbf{\tau}}^\intercal\cdot{\mathbf{W}}_{\mu}\right)-m\right]\uppsi^{{\rm c}}=0. \label{succ}
\end{eqnarray}
In order to implement a covariant gauge potential,    Eq. (\ref{succ}) can be thus multiplied by $i\tau_2$, together with the  Pauli's identity $\tau_{a}=-\tau_{2} \tau_{a}^\intercal\tau_{2}^{-1}$, to yield $
\left[\slash\!\!\!\partial-\frac{g}{2}{\mathbf{\tau}}\cdot\slashed{\mathbf{W}}-m\right]\tilde\uppsi=0$, 
where $\tilde\uppsi\equiv i\tau_{2}\uppsi^{{\rm c}}$ denotes the isospin-charge   conjugate  spinor \cite{Inglis:2012rg}. Defining
 \begin{eqnarray}
\Omega =\tau^{a}\upgamma^{\mu}\uppsi W_{a\mu}=  \frac2g(\slash\!\!\!\partial-m)\uppsi, \label{deriv}
\end{eqnarray}
 and multiplying the first equation in (\ref{deriv}) by $\bar\uppsi\tau_{a}\upgamma_{\mu}$, it reads $
\tau_{a}\tau^{b}W_{b\mu}=(\delta_{ab}+i\,\epsilon_{a}{}^{bc}\tau_{c})W_{b\mu},$ yielding therefore \begin{eqnarray}\label{phi12}
\bar\uppsi(\delta_{\mu}{}^{\nu}-i\upgamma_{\mu}{}^{\nu})\uppsi W_{a\nu}+\epsilon_{abc}\bar\uppsi\tau_{c}(i\delta_{\mu}{}^{\nu}+\upgamma_{\mu}{}^{\nu})\uppsi W_{b\nu}=\bar\uppsi\tau_{a}\upgamma_{\mu}\Omega.
\end{eqnarray}
The analogue non-Abelian bispinors are, then, defined by \cite{Booth:2001ey,Inglis:2012rg} 
\begin{equation}\label{nonabili}
\bar{\tilde\uppsi}(\tau_{i}\otimes\Gamma)\tilde\uppsi=-\bar\uppsi(\tau_2^{-1}\tau_{i}{}^\intercal\tau_2)\otimes(C^{-1}\Gamma^\intercal C)\uppsi,
\end{equation}
where $\Gamma$ is an arbitrary multivector in the Clifford-Dirac spacetime algebra and the (Euclidean) indexes run as $i=1,2,3$, for $i=0$ corresponding to the $2\times2$ identity. Eq. (\ref{phi12})  implies that 
\begin{equation}\label{equazione1}
gW_{a\mu}\bar\uppsi\uppsi + g\epsilon_{abc}W_{b}{}^{\nu}\bar\uppsi\tau^{c}\upgamma_{\mu\nu}\uppsi=i\,\bar\uppsi(\tau_{a}\upgamma_{\mu}\slashed{\partial}-\overleftarrow{\slashed{\partial}}\tau_{a}\upgamma_{\mu})\uppsi-2m\bar\uppsi\tau_{a}\upgamma_{\mu}\uppsi.
\end{equation}

The non-Abelian bilinear covariants are defined by ($i=0,1,2,3$): 
\begin{subequations}
\begin{eqnarray}
\sigma_{i}&=&\bar\uppsi\tau_{i}\uppsi, \label{bt1}\\
J_{i\mu}&=&\bar\uppsi\tau_{i}\upgamma_{\mu}\uppsi, \label{bt2}\\
S_{i\mu\nu}&=&\bar\uppsi\tau_{i}\upgamma_{\mu\nu}\uppsi, \\
K_{i\mu}&=&\bar\uppsi\tau_{i}\upgamma_{5}\upgamma_{\mu}\uppsi, \label{bt4}\\
\omega_{i}&=&\bar\uppsi\tau_{i}\upgamma_{5}\uppsi,\label{bt5}
\end{eqnarray}
\end{subequations}
{\color{black}{originating the  classification of non-Abelian spinor fields  into the following disjoint classes ($i=0,1,2,3$; $j=1,2,3$):
\begin{subequations}
\begin{eqnarray}
&&1)\;\sigma\neq0,\;\;\;\sigma_j\neq0,\;\;\;\omega\neq 0,\;\;\;\omega_j\neq 0,\;\;\;\mathit{K}_{i\mu}\neq0, \;\;\;\mathit{S}_{i\mu\nu}\neq0\;\; \label{stipo1}\\  
&&2)\;\sigma=0,\;\;\;\sigma_j\neq0,\;\;\;\omega\neq 0,\;\;\;\omega_j\neq 0,\;\;\;\mathit{K}_{i\mu}\neq0, \;\;\;\mathit{S}_{i\mu\nu}\neq0\;\; \\  
&&3)\;\sigma\neq0,\;\;\;\sigma_j=0,\;\;\;\omega\neq 0,\;\;\;\omega_j\neq 0,\;\;\;\mathit{K}_{i\mu}\neq0, \;\;\;\mathit{S}_{i\mu\nu}\neq0\;\; \\  
&&4)\;\sigma=0,\;\;\;\sigma_j=0,\;\;\;\omega\neq 0,\;\;\;\omega_j\neq 0,\;\;\;\mathit{K}_{i\mu}\neq0, \;\;\;\mathit{S}_{i\mu\nu}\neq0\;\; \\   
&&5)\;\sigma\neq0,\;\;\;\sigma_j\neq0,\;\;\;\omega\neq 0,\;\;\;\omega_j= 0,\;\;\;\mathit{K}_{i\mu}\neq0, \;\;\;\mathit{S}_{i\mu\nu}\neq0\;\; \\  
&&6)\;\sigma\neq0,\;\;\;\sigma_j\neq0,\;\;\;\omega= 0,\;\;\;\omega_j\neq 0,\;\;\;\mathit{K}_{i\mu}\neq0, \;\;\;\mathit{S}_{i\mu\nu}\neq0\;\; \\  
&&7)\;\sigma\neq0,\;\;\;\sigma_j\neq0,\;\;\;\omega= 0,\;\;\;\omega_j= 0,\;\;\;\mathit{K}_{i\mu}\neq0, \;\;\;\mathit{S}_{i\mu\nu}\neq0\;\; \\  
&&8)\;\sigma=0,\;\;\;\sigma_j=0,\;\;\;\omega= 0,\;\;\;\omega_j= 0,\;\;\;\mathit{K}_{i\mu}\neq0, \;\;\;\mathit{S}_{i\mu\nu}\neq0\;\; \label{class8}\\   
&&9)\;\sigma=0,\;\;\;\sigma_j=0,\;\;\;\omega= 0,\;\;\;\omega_j= 0,\;\;\;\mathit{K}_{0\mu}=0,\;\;\;\mathit{K}_{j\mu}\neq0, \;\;\;\mathit{S}_{i\mu\nu}\neq0\;\; \\   
&&10)\;\sigma=0,\;\;\;\sigma_j=0,\;\;\;\omega= 0,\;\;\;\omega_j= 0,\;\;\;\mathit{K}_{i\mu}\neq0,\;\;\;\mathit{S}_{0\mu\nu}=0,\;\;\;\mathit{S}_{j\mu\nu}\neq0\;\; \\   
&&11)\;\sigma=0,\;\;\;\sigma_j=0,\;\;\;\omega= 0,\;\;\;\omega_j= 0,\;\;\;\mathit{K}_{i\mu}=0, \;\;\;\mathit{S}_{i\mu\nu}\neq0\;\; \\   
&&12)\;\sigma=0,\;\;\;\sigma_j=0,\;\;\;\omega= 0,\;\;\;\omega_j= 0,\;\;\;\mathit{K}_{i\mu}=0,\;\;\;\mathit{S}_{0\mu\nu}=0,\;\;\;\mathit{S}_{j\mu\nu}\neq0\\
&&13)\;\sigma=0,\;\;\;\sigma_j=0,\;\;\;\omega= 0,\;\;\;\omega_j= 0,\;\;\;\mathit{K}_{i\mu}\neq0, \;\;\;\mathit{S}_{i\mu\nu}=0\\
&&14)\;\sigma=0,\;\;\;\sigma_j=0,\;\;\;\omega= 0,\;\;\;\omega_j= 0,\;\;\;\mathit{K}_{0\mu}=0,\;\;\;\mathit{K}_{j\mu}\neq0, \;\;\;\mathit{S}_{i\mu\nu}=0.\;\;\label{stipo10} \end{eqnarray}
\end{subequations}
The classes 1) -- 7) correspond to SU(2) $\times$ U(1) regular spinors, whereas classes 8) -- 14) are SU(2) $\times$ U(1) singular spinors. It is worth to mention that 
the field ${\mathfrak{J}} = J_{i\mu}\upgamma^\mu\tau_i$ can be thought as an SU(2) $\times$ U(1) current density, for the classes 1) -- 7) of regular spinors. Analogously to the previous geometric interpretation to singular spinors in the Lounesto's classification, class 8) consists of SU(2) $\times$ U(1) flag-dipole spinors. Up to the class 8), all  spinor classes are in close straightforward non-Abelian generalizations. However, new aspects are unveiled with non analogy to the Lounesto's classification,  from class 9) to class 14). In fact, class 9) correspond to SU(2)  flagdipole-U(1) flagpole spinors. This feature is completely unexpected, as the bispinors in this class are flag-dipole  spinors with respect to the SU(2) sector ($j=1,2,3$), having 4 non-null flags, $S_{i\mu\nu}$ and 3 non-Abelian poles, $K_{j\mu}$; however, 1 pole given by $K_{0\mu}\equiv K_\mu$, corresponding to the U(1) sector ($i=0$) given by the usual bilinear covariant $K_\mu=\bar\psi\gamma_\mu\gamma_5\psi$, equals zero.   Class 10) is characterized 
by SU(2)  flagpole-U(1) dipole spinors. Indeed, the bispinors in class 10) are flagpole  spinors with respect to the SU(2) sector ($j=1,2,3$), having 4 non-null poles, $K_{i\mu\nu}$ and 3 non-Abelian flags, $S_{j\mu\nu}$, with 1 additional flag  $S_{0\mu\nu}\equiv S_{\mu\nu}$, corresponding to the U(1) sector ($i=0$) given by the usual bilinear covariant $S_{\mu\nu}=\bar\psi\gamma_{\mu\nu}\gamma_5\psi$, that is equal to  zero.   Class 11) is a case of a pure class of SU(2) flagpoles, as well as class 13) is also a pure class of SU(2) dipoles. On the other hand, the class 12) consist of SU(2) flagpoles-U(1) poles and class 14) consist of SU(2) dipoles-U(1) poles. Besides, the inherent geometric
structure underlying the bispinor classes (\ref{stipo1} -- \ref{stipo10}) relies on the existence of four flags, $S_{i\mu\nu}$, and eight poles, $K_{i\mu}$ and $J_{i\mu}$, in the defining Eqs.  (\ref{bt2} -- \ref{bt4}), corresponding to a 4-fold richer structure than the one provided by the Lounesto's classification. Moreover, classes 12) and 14) admits, respectively,  subclasses of SU(2) flagpoles and dipoles bispinors, that are U(1) poles. This has non analogy to the Lounesto's classification, having for the classes 12) and 14) all the bilinears, but {\bf J}, vanishing. }}

In order to emulate the (ghost) spinors that extend the Lounesto's classification in Ref. \cite{EPJC}, we present the following  additional  spinors classes:
\begin{subequations}
\begin{eqnarray}
&&15)\;\sigma=0,\;\;\;\sigma_j=0,\;\;\;\omega= 0,\;\;\;\omega_j= 0,\;\;\;\mathit{J}_{i\mu}=0,\;\;\;\mathit{K}_{i\mu}\neq0, \;\;\;\mathit{S}_{i\mu\nu}\neq0\;\; \\   
&&16)\;\sigma=0,\;\;\;\sigma_j=0,\;\;\;\omega= 0,\;\;\;\omega_j= 0,\;\;\;\mathit{J}_{i\mu}=0,\;\;\;\mathit{K}_{i\mu}=0, \;\;\;\mathit{S}_{i\mu\nu}\neq0\;\; \\   
&&17)\;\sigma=0,\;\;\;\sigma_j=0,\;\;\;\omega= 0,\;\;\;\omega_j= 0,\;\;\;\mathit{J}_{i\mu}=0,\;\;\;\mathit{K}_{i\mu}\neq0, \;\;\;\mathit{S}_{i\mu\nu}=0.\;\;
\end{eqnarray}
\end{subequations}
It is immediate to notice that the Lounesto's classification arises as a Pauli's singlet corresponding to $i=0$ in the classification (\ref{stipo1} -- \ref{stipo10}). In fact, when $i=0$,  Eq. (\ref{bt1}) reads $\sigma_0 = \bar\uppsi\tau_0\uppsi=\bar\uppsi I\uppsi = \bar\uppsi\uppsi = \sigma$. When $i=0$, the other SU(2) $\times$ U(1) bilinears (\ref{bt2} -- \ref{bt5}) are also led to their usual U(1) bilinears, namely, $J_{0\mu}\mapsto J_\mu$, $S_{0\mu\nu} \mapsto S_{\mu\nu}$, $K_{0\mu}\mapsto K_\mu$ and $\omega_0=\omega$. 
Hence, when $i=0$, the above bilinears correspond to a Pauli's  
singlet, and the (isomorphic) Lounesto's classification arises, as $\sigma_0=\bar\uppsi\tau_{0}\uppsi=\sigma\otimes I$, which vanished if $\sigma$ equals zero.
A similar analysis holds for $\omega_{0}=\bar\uppsi\tau_{0}\upgamma_{5}\uppsi = \omega\otimes I$.

With the definition of the non-Abelian bilinear covariants, Eq. (\ref{equazione1}) can now be rewritten in a more condensed form \cite{Inglis:2012rg}
\begin{equation}\label{eq345}
\left(\delta_{\mu}{}^{\nu}\delta_{a}{}^{b}J_{0}-S_{c\mu}{}^{\nu}\epsilon_{a}{}^{cb}\right)gW_{b\nu}=i\uppsi(\tau_{a}\upgamma_{\mu}\slashed{\partial}-\overleftarrow{\slashed{\partial}}\tau_{a}\upgamma_{\mu})\uppsi-2mJ_{a\mu}. 
\end{equation}
 Now, defining $\star S_{i\mu\nu}=\frac{i}{2}\epsilon_{\mu\nu\alpha\beta}S_{i}^{\;\,\alpha\beta}$ and  multiplying Eq. (\ref{deriv}) by $\bar\uppsi\tau_{a}\upgamma_{5}\upgamma_{\mu}$ yields 
\begin{equation}\label{inversione}
\left(\delta_{\mu}{}^{\nu}\delta_{a}{}^{b}K_{0}-\star S_{c\mu}{}^{\nu}\epsilon_{a}{}^{cb}\right)gW_{b\nu}=i\,\bar\uppsi(\tau_{a}\upgamma_{5}\upgamma_{\mu}\slashed{\partial}+\overleftarrow{\slashed{\partial}}\tau_{a}\upgamma_{5}\upgamma_{\mu})\uppsi.
\end{equation}
Adding Eqs. (\ref{eq345}) and (\ref{inversione}) implies that 
\begin{eqnarray}\label{wminus1}
\left[\epsilon_{a}{}^{cb}(\star S_{c\mu}{}^{\rho}+S_{c\mu}{}^{\rho})-\delta_{\mu}{}^{\rho}\delta_{a}{}^{b}{(J_{0}+K_{0})}\right]W_{b\rho} \nonumber \\
=-\frac{i}{g}{[\bar\uppsi\tau_{a}\upgamma_{\mu}(I+\upgamma_5)\slashed{\partial}\uppsi-\bar\uppsi\overleftarrow{\slashed{\partial}}\tau_{a}\upgamma_{\mu}(I-\upgamma_5)\uppsi]-2m({J_{0}+K_{0}})J_{a\mu}},
\end{eqnarray}
Since the left-hand side of the above equations is invertible, a Neumann series analysis 
implies that \cite{Inglis:2012rg}
\begin{equation}
\frac{g}{2}J^{a\nu}W_{a\nu}=i\,\bar\uppsi\slash\!\!\!\partial\uppsi-mJ_{0},
\end{equation}
explicitly providing the coupling between the Lorentz non-Abelian density current $J_{a\mu}$ and the vector potential field. It is worth to emphasize that interpreting 
the $J_{a\mu}$ as a non-Abelian density current holds for non-Abelian regular spinors in classes 1) -- 7). 
In the next section, the generalized Fierz identities are briefly reviewed and introduced. 

\section{Non-abelian Fierz aggregate and Fierz identities}
Now, the non-Abelian analogs of Eqs. (\ref{fie1} -- \ref{FIERZ}) can be now studied, considering  $\uppsi\bar\uppsi$ $8\times8$ matrices. We have already seen that  Eq. (\ref{zsigma}) represents the Fierz aggregate. The non-Abelian bilinear covariants, hence, 
make the definition of the non-Abelian Fierz aggregate 
\begin{eqnarray}\label{nafierz}
Z_\uppsi=\omega_{i}(\upgamma_{5}\otimes\tau^{i})-K_{i\mu}(\upgamma_{5}\upgamma^{\mu}\otimes\tau^{i})
+S_{i\mu\nu}(\gamma^{\mu\nu}\otimes\tau^{i})
+J_{i\mu}(\upgamma^{\mu}\otimes\tau^{i}) 
+\sigma_{i}(I\otimes\tau^{i}), 
\end{eqnarray}
where the coefficients are the non-Abelian bilinear covariants (\ref{bt1} -- \ref{bt5}), consisting of SU(2)  bispinors. 

Hence, the Fierz identities (\ref{fie1} -- \ref{FIERZ}) can be then generalized for the non-Abelian case, yielding, for example the following expression \cite{Inglis:2012rg} (hereon the expressions for symmetrized [antisymmetrized] indexes 
$A_{(\mu\nu)}=A_{\mu\nu}+A_{\nu\mu} \;\;[A_{[\mu\nu]}=A_{\mu\nu}-A_{\nu\mu}$], for any tensor $A_{\mu\nu}$ and higher order generalizations, shall be used):
\begin{eqnarray}
{} J_{a}{}^{\mu}K_{b}{}^{\nu}&=&\bar\uppsi\tau_{a}\upgamma^{\mu}\uppsi\bar\uppsi\tau_{b}\upgamma_{5}\upgamma^{\nu}\uppsi \nonumber \\
&=&\frac14\left[i\, J_{(a}\star S_{b)}{}^{\mu\nu}-i\, K_{(a}S_{b)}{}^{\mu\nu}+J_{(a}{}^{(\mu}K_{b)}{}^{\nu)}-J_{c}{}^{(\mu}K^{c\nu)}+\delta_{ab}\left(-J_{0\sigma}K_{0}{}^{\sigma}+J_{c\sigma}K^{c\sigma}\right)\eta^{\mu\nu}\right.\nonumber \\
&&\left.+\delta_{ab}(i\, J_{0}\star S_{0}{}^{\mu\nu}-i\, J_{c}\star S^{c\mu\nu}-i\, K_{0}S_{0}{}^{\mu\nu}+i\, K_{c}S^{c\mu\nu}+J_{0}{}^{(\mu}K_{0}{}^{\nu)})\right] \nonumber \\
&&+\frac14\epsilon_{ab}{}^{c}\left[i (K_{0}J_{c}+ J_{0}K_{c})\eta^{\mu\nu}+(J_{c\sigma}J_{0\lambda}+K_{c\sigma}K_{0\lambda})\epsilon^{\mu\nu\sigma\lambda}-\frac{i}{2}S_{(0}{}^{(\mu}{}_{\vert\sigma\vert}\star S_{c)}{}^{\vert\sigma\vert\nu)}\right].
\end{eqnarray}

Emulating the Fierz identity 
$
S^{\mu\nu}=\frac{1}{\sigma^{2}-\omega^{2}}[\sigma\epsilon^{\mu\nu}_{\;\;\;\rho\chi}-i\,\omega\epsilon_{\alpha\rho\chi}\epsilon^{\alpha\mu\nu}]J^{\rho}K^{\chi}, 
$ \cite{Cra,Mosna:2003am} for the $i=0$ case that corresponds to a Pauli's singlet  equivalent to the Lounesto's classification,   one can further calculate other Fierz identities for the non-Abelian case, as 
\begin{equation}
J_{i}{}^{[\mu}K^{i\nu]}=2i\,(J_{0}\star S_{0}{}^{\mu\nu}-K_{0}S_{0}{}^{\mu\nu}).
\end{equation}
Hence, adding the term $\epsilon^{\mu\nu\rho\chi}J_{0\rho}K_{0\chi}$ to $\epsilon^{\mu\nu\alpha\beta}J_{a\alpha}K^{a}{}_{\beta}$ yields \cite{Inglis:2012rg} 
\begin{equation}
\epsilon^{\mu\nu\alpha\beta}J_{i\alpha}K^{i}{}_{\beta}=2(J_{0}S_{0}{}^{\mu\nu}-K_{0}\star S_{0}{}^{\mu\nu}),
\end{equation}
following that 
\begin{equation}\label{fierz156000}
S_{0}{}^{\mu\nu}=\frac{1}{2(J_{0}^{2}-K_{0}^{2})}\left(J_{0}\epsilon^{\mu\nu}{}_{\rho\chi}-i\, K_{0}\epsilon_{\alpha\rho\chi}\epsilon^{\alpha\mu\nu}\right)J_{i}{}^{\rho}K^{i\chi}.
\end{equation}
Moreover a generalized Fierz identity holds for the 
non-Abelian density, written as a function of the non-Abelian chiral current and the non-Abelian spin density \cite{Inglis:2012rg}:
\begin{eqnarray}\label{fierz156}
{} S_{a}{}^{\mu\nu}&=&\frac{J_{(a}^{\;\rho}K_{0)}^{\;\chi}}{J_{0}^{2}\!-\!K_{0}^{2}}\left[J_{0}\epsilon^{\mu\nu}_{\;\;\rho\chi}\!-\!i K_{0}\epsilon_{\alpha\rho\chi}\epsilon^{\alpha\mu\nu}\right]
\!-\!\frac{J_{0}^{2}\!+\!K_{0}^{2}}{2(J_{0}^{2}\!-\!K_{0}^{2})^{2}}\left[J_{a}\epsilon^{\mu\nu}{}_{\rho\chi}\!+\!i K_{a}\epsilon_{\alpha\rho\chi}\epsilon^{\alpha\mu\nu}\right]J_{i}{}^{\rho}K^{i\chi} \nonumber \\
&&\qquad\qquad\qquad+\frac{J_{0}K_{0}}{(J_{0}^{2}-K_{0}^{2})^{2}}\left[K_{a}\epsilon^{\mu\nu}{}_{\rho\chi}+i\, J_{a}\epsilon_{\alpha\rho\chi}\epsilon^{\alpha\mu\nu}\right]J_{i}{}^{\rho}K^{i\chi}.
\end{eqnarray}

\section{Conclusions}
Due to the limitations of the Lounesto's spinor field  classification we have proposed an extended non-Abelian spinor field  classification that encompasses the 
SU(2) $\times$ U(1) gauge symmetries, responsible for the conservation of the electroweak and electromagnetic conserved charges, by the Noether's theorem.
This generalized spinor field classification can be still led to the Lounesto's classification, considering the identity 
2 $\times$ 2 matrix ($i=0$) into all the expressions in Sect. 3. In particular, the U(1) gauge bilinear covariants, that compose to the original Lounesto's classification, 
are obtained as the particular case of a  Pauli's singlet, in the non-Abelian spinor field classification. Non-Abelian generalized Fierz aggregates and some of the corresponding non-Abelian generalized Fierz identities have been also studied. 

Although the SU(2) $\times$ U(1) gauge symmetry was chosen to be the fundamental gauge symmetry to illustrate the 14 new classes of regular and singular non-Abelian spinors in Eqs. (\ref{stipo1} -- \ref{stipo10}), 
SU(3) gauge symmetries can be analogously introduced, 
with the immediate difference that the gauge indexes 
should run as $i=1,\ldots,8$, being again $i=0$
correspondent to the Lounesto's classification.
Obviously, the similar generalized Fierz identities of Sect. IV should be derived for the SU(3) gauge symmetric case, which is not our current goal here. {\color{black}{In fact, any gauge group  $G$, with associated Lie algebra $\mathfrak{g}$, can be used for an immediate generalization of the non-Abelian bilinear covariants and the classification (\ref{stipo1} -- \ref{stipo10}), when one considers a set  $\{\uptau_{a}\}_{a=1}^{{\rm rank}\; G}$ of generators, satisfying  $[{\uptau_{a}}, {\tau_{b}}]=f_{ab}{}^{c}{\tau_{c}}$. }}

\acknowledgments
RdR~is grateful to CNPq (Grant No. 303293/2015-2),
and to FAPESP (Grant No. 2017/18897-8) and to INFN, for partial financial support. 
\end{document}